\documentclass[%
 reprint,
nofootinbib,
 amsmath,amssymb,
 aps,
]{revtex4-1}

\usepackage[pdftex]{graphicx}
\usepackage{dcolumn}
\usepackage{bm}
\usepackage{subfigure}

\begin{document}


\title{Local gravity test of unified models of inflation and dark energy\\
in $f(R)$ gravity}

\author{Mai Yashiki}
 \email{g005wb@yamaguchi-u.ac.jp}
\author{Nobuyuki Sakai}
 \email{nsakai@yamaguchi-u.ac.jp}
\author{Ryo Saito}
\email{rsaito@yamaguchi-u.ac.jp}
\affiliation{Graduate School of Sciences and Technology for Innovation,
Yamaguchi University, Yamaguchi 753-8512, Japan}


\date{\today}

\begin{abstract}

We consider $f(R)$ gravity theories which unify $R^n$ inflation and dark energy models. First, from the final Planck data of the cosmic microwave background, we obtain a condition, $1.977 < n < 2.003$. Next, under this constraint, we investigate local-gravity tests for three models. We find that the $R^n$ term can dominate over the dark energy term even at the Earth's curvature scale, contrary to intuition; however, the $R^n$ term does not relax or tighten the constraints on the three models.

\end{abstract}

\pacs{Valid PACS appear here}
\maketitle


\section{\label{sec:intro}Introduction}

There have been growing evidences that the Universe experiences two accelerated expansion eras: inflation \cite{guth,sato,planck2018_X} and the late-time acceleration \cite{1a,1a2,1a3}. 
However, in spite of many attempts, the origin(s) of both cosmic accelerations has/have not been identified yet.

Both for inflation and the late-time acceleration, models are classified into two categories: dark energy (DE) and modified gravity (MG).
In the first approach,  one introduces exotic matter components such as the cosmological constant or a scalar field,
which is called DE for the late-time acceleration. 
If the exotic component has an equation of state 
with $w \equiv P/\rho < -1/3$, 
it drives the accelerated expansion of the Universe.
The other approach is to modify the theory of gravity from general relativity (GR) \cite{mog,mog2,mog3}.
A simple family of the modified gravity is $f(R)$ theories \cite{barrow,maeda}, where the Ricci scalar $R$ in the Einstein-Hilbert action is replaced by a nonlinear function $f(R)$.
In $f(R)$ gravity, by choosing an appropriate function $f(R)$, 
the expansion of the Universe is accelerated without introducing any exotic matter components.

An example of an inflationary model in $f(R)$ theories is the Starobinsky model $f(R)=R+\alpha R^2$ \cite{tomita,r^2}.
The Starobinsky model is known as the first inflationary model and its prediction gives a good fit to the Planck data for the spectral index $n_s$ and the tensor-to-scalar ratio $r$ \cite{inf_bestmodel, planck2018_X}.
On the other hand, 
it is known that some other forms of $f(R)$ can drive the late-time acceleration \cite{fr-DE, plmodel, fr-grav, AB-Do}. 

On this ground, 
it has been considered whether a single $f(R)$ model can describe both the early and late accelerated expansion eras in a unified way \cite{ABS,0608008,unif_model, lalak}. 
Such unified models should not only be viable as a model for the two accelerations but also satisfy a local-gravity test.
In $f(R)$ theories, 
the law of gravity in a local system is modified as well as that on cosmological scales.
In more words, 
there is an extra scalar degree of freedom with a universal coupling to matter, which is called scalaron \cite{r^2}.
Therefore, it mediates a fifth force and can violate local-gravity constraints.  
A possible resolution for this problem is given by the chameleon screening mechanism \cite{chameleon, chameleon2}, 
which makes the range of the fifth force short in high-density regions and an $f(R)$ model compatible with the local-gravity constraints.
In fact, for some $f(R)$ models of the late-time acceleration \cite{HuSawicki,StaDE}, the chameleon screening mechanism is known to work without spoiling their success in cosmology.

In this paper we consider a class of $f(R)$ unified models:
\begin{equation}\label{fR}
f(R) = R+\alpha R^n + f_{\rm DE}(R) \,,
\end{equation}
and give local-gravity constraints on three concrete models. 
In previous studies \cite{lalak, 0707.1941, 0710.1738, 0712.4017, 1012.2280} there have been several discussions related to this subject.
Artymowski and Lalak \cite{lalak} studied the model of $f_{\rm DE}(R)= -\beta R^{2-n}$ and claimed that their unified model is consistent with local-gravity tests due to the existence of $\alpha R^n$ term (hereafter, the $\alpha$ term), 
though they did not give any concrete analysis on the fifth force. 
One of our motivations is to verify their conclusion.

Nojiri {\it et al.} \cite{0707.1941, 0712.4017, 1012.2280} studied some unified models where the inflation term approaches constant as $R$ goes to infinity, which is a different class of models from (\ref{fR}). Nojiri and Odintsov \cite{0710.1738} also investigated a model of $f(R)=R+(\alpha R^{2n}-\beta R^n)/(1+\gamma R^n)$. 
This model does not fall into the class (\ref{fR}) in a strict sense; however, their inflationary term approaches $\propto R^n$ as $R$ becomes large enough. In this sense our work is an extension of Ref. \cite{0710.1738}.

In the local-gravity analysis, we newly investigate two points: (i) how important the inflationary term can be and (ii) the full analysis of the fifth force taking into account the chameleon screening mechanism in the unified models. 
In the above models \cite{lalak, 0710.1738}, 
the inflationary and DE terms are related and hence it is difficult to discuss how the inflationary term affects the local-gravity analysis. 
To avoid this subtlety, 
we consider three unified models where the model parameters in the inflationary and DE terms are independent. 
One may think intuitively that the inflationary term is negligible in local-gravity analysis. 
However, we find that the inflationary term can dominate over the DE term even at the Earth's curvature scale, 
depending on model parameters. 
We therefore stress that the inflationary term should also be taken into account to estimate the fifth force and to give constraints models from local-gravity tests. 
Then, we give the full analysis of the fifth force in the unified models without dropping the inflationary term. 
The scalaron's mass has been estimated in the literature but we also estimate the scalar charge, or the thin-shell parameter, of a gravitational object. 
To determine the scalar charge, one needs to know the global field profile of the scalaron from the inner region to the outer region of the object, 
where the curvature scale varies from the astrophysical scale to the cosmological scale. 
Therefore, 
both of the inflationary and DE terms can be relevant in this analysis.

The unified models should consistently describe the evolution of the Universe from inflation to the current accelerated expansion. 
Therefore, 
the model parameters in the inflationary term should be constrained from 
the inflationary era, 
i.e. the constraints from the primordial spectral index $n_s$ and the tensor-to-scalar ratio $r$. 
Then, under this constraint, 
we analyze the chameleon screening mechanism with the inflationary term for the three unified models.

This paper is organized as follows. 
First, in Sec. \ref{sec:f(R)grav}, we show basic equations of $f(R)$ theories used in this paper.
In Sec. \ref{sec:model}, we introduce the unified models and their corresponding form in the Einstein frame.
In Sec. \ref{sec:inflation}, 
the inflationary constraints are derived for the unified models, which is extended from our previous paper \cite{my1st}.
In Sec. \ref{sec:gravity test}, the local-gravity test is studied for the unified models.
Finally, Sec. \ref{sec:conclusion} is devoted to the conclusion.

In the following, we set $8\pi G \equiv M_{\rm pl}^{-2} =1$, where $G$ is the gravitational constant and $M_{\rm pl}$ is the reduced Planck mass.

\section{\label{sec:f(R)grav} Basic equations}

To begin with, let us briefly show some basic equations of a general $f(R)$ model used in this paper. 
(See, e.g., Ref. \cite{fr-grav} for a more extensive review.)

The action of $f(R)$ gravity is given by
\begin{equation}
{\cal S}=\frac{1}{2} \int d^4 x \sqrt{-g} f(R) +\int d^4 x \sqrt{-g} {\cal L}_{\rm m} (g_{\mu \nu}, \Phi_{\rm m}),
\label{eq:fr action jordan}
\end{equation}
where ${\cal L}_{\rm m}$ is a matter Lagrangian density and a matter field $\Phi_m$ is assumed to minimally couple to the metric $g_{\mu\nu}$.

Varying Eq. (\ref{eq:fr action jordan}) with respect to $g_{\mu\nu}$, we obtain
\begin{eqnarray}
F(R)R_{\mu \nu}(g) &-& \frac{1}{2}f(R)g_{\mu \nu}-\nabla_\mu \nabla_\nu F(R) \nonumber \\
&+& g_{\mu \nu} \Box F(R)= T^{({\rm m})}_{\mu \nu},
\label{eq:fr action variation}
\end{eqnarray}
where $F(R)\equiv {\rm d}f/{\rm d} R$ and $T^{({\rm m})}_{\mu\nu}$ is the energy-momentum tensor of the matter field.
The trace of Eq. (\ref{eq:fr action variation}) gives
\begin{equation}
3 \Box F +FR-2f= T^{({\rm m})},
\label{eq:motion eq jordan}
\end{equation}
where $T^{({\rm m})} \equiv g^{\mu\nu}T^{({\rm m})}_{\mu\nu}$. 
This propagating degree of freedom $F$ corresponds to the scalaron, which is absent in general relativity because it corresponds to $F = 1$. 
\\

It is sometimes more intuitive to use the Einstein frame action as follows \cite{e-frame,e-frame2}.
The action (\ref{eq:fr action jordan}) can be rewritten as
\begin{equation}
{\cal S}= \int d^4 x \sqrt{-g} \left( \frac{1}{2} FR -U(F) \right) +\int d^4 x {\cal L}_{\rm m} (g_{\mu \nu}, \Phi_{\rm m}),
\label{eq:fr action jordan rewrite}
\end{equation}
where $U(F)$ is defined by
\begin{equation}
U(F)=\frac{FR-f}{2}.
\label{eq:potential jordan}
\end{equation}
To obtain the action in the Einstein frame, 
we make the conformal transformation \cite{e-frame,maeda},
\begin{equation}
\tilde{g}_{\mu \nu}=F g_{\mu \nu},
\label{eq:conformal transformation}
\end{equation}
where a tilde denotes a quantity in the Einstein frame.
Here, we introduce a scalar field $\phi$ by
\begin{equation}
\phi \equiv \sqrt{\frac{3}{2}} \ln{F}.
\label{eq:phi-F}
\end{equation}
Using Eqs. (\ref{eq:conformal transformation}) and (\ref{eq:phi-F}), the action (\ref{eq:fr action jordan rewrite}) can be rewritten as
\begin{equation}
\begin{split}
{\cal S}_{\rm E} &= \int d^4 x \sqrt{-\tilde{g}} \left[ \frac{1}{2} \tilde{R}-\frac{1}{2} \tilde{g}^{\mu \nu} \partial_\mu \phi \partial_\nu \phi -V(\phi) \right] \\
&\qquad +\int d^4 x {\cal L}_{\rm m} (F^{-1}(\phi) \tilde{g}_{\mu \nu}, \Phi_{\rm m}),
\label{eq:fr action einstein}
\end{split}
\end{equation}
where $V(\phi)$ is the field potential in the Einstein frame:
\begin{equation}
V(\phi)=\frac{U}{F^2} =\frac{FR-f}{2 F^2}
\label{eq:potential einstein}.
\end{equation}
In the Einstein frame, the scalar field equation is given by
\begin{equation}
\widetilde{\square} \phi = \frac{{\rm d} V_{\rm eff}}{{\rm d} \phi}\,,
\label{eq:phi eom cov}
\end{equation}
where the effective potential is defined as
\begin{equation}
V_{\rm eff} (\phi)= V(\phi) + \rho^* e^{-\frac{\phi}{\sqrt{6}}}\,,
\label{eq:V_eff}
\end{equation}
for a non-relativistic object \cite{chameleon}. 
Here, $\rho^*$ is a conserved quantity in the Einstein frame, which is related to the energy density $\rho$ in the Jordan frame as $\rho^* = e^{-3\phi / \sqrt{6}} \rho$. 
In relevant cases, the difference between these two densities is negligible because $\phi \ll 1$. 
Therefore, we will omit the asterisk hereafter.

\section{\label{sec:model} $f(R)$ unified models}

We consider a class of $f(R)$ unified models (\ref{fR}). 
The $f(R)$ models should satisfy viable conditions that tachyonic and ghost instabilities are absent. The conditions can be written in terms of $F$ as
\begin{equation}
F >0  \\\ ,  \hspace{0.5cm}  F_{,R} \equiv {\rm d}F/{\rm d}R >0 \\\ ,  \hspace{0.5cm}  {\rm for} \hspace{0.5cm} R \geq R_0,
\label{eq:stability conditions}
\end{equation}
where $R_0$ is the present value of the Ricci scalar. (See, e.g., Ref. \cite{fr-grav} for more details on these issues.)

We consider three types of the DE terms, $f_{\rm DE}(R)$, as follows.

\subsection{Model 1: Power-law DE model}

First, we introduce the following unified model:
\begin{equation}
f(R) = R+\alpha R^n - \beta R^p \,,
\label{eq:ex lalak}
\end{equation}
with $n>1$, $0< p < 1$ and $\alpha, \beta>0$.
This model is an extension of the model proposed by Artymowski and Lalak \cite{lalak}, $f(R)=R+\alpha R^n -\beta R^{2-n}$.
For $\alpha = 0$, this model corresponds to the power-law DE model \cite{fr-DE,plmodel}, which satisfies the conditions (\ref{eq:stability conditions}).

In the model (\ref{eq:ex lalak}), 
the field potential in the Einstein frame (\ref{eq:potential einstein}) is read as
\begin{equation}
V(\phi)=\frac{e^{-\frac{4}{\sqrt{6}}\phi}}{2} \left[ \alpha (n-1) R(\phi)^n + \beta (1-p) R(\phi)^p \right] \,.
\label{eq:potentail_ab}
\end{equation}
Here, $R(\phi)$ is obtained by solving the equation,
\begin{equation}
 e^{\frac{2}{\sqrt{6}}\phi} = F(R) = 1 + \alpha nR^{n-1} - \beta p R^{p-1}\,,
 \label{eq:phi r}
\end{equation}
which is regular 
when the stability conditions (\ref{eq:stability conditions}) are satisfied. 

\begin{figure}[t]
\includegraphics[width=8.5cm]{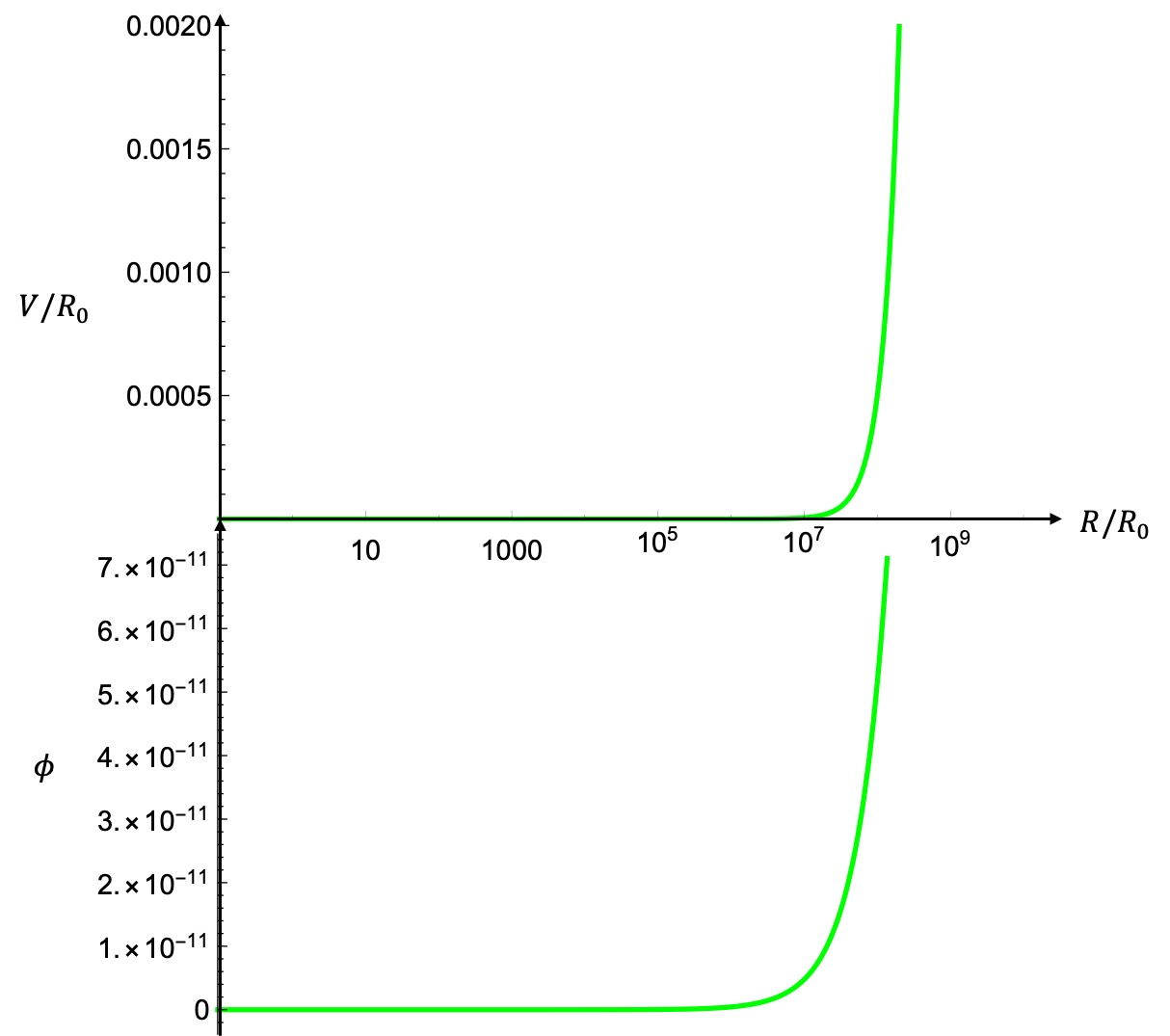}
\caption{Potential in the Einstein frame for $R^2$ inflation model. 
We show parametric representations $V(R)$ and $\phi(R)$ in the top and the bottom, respectively.
In this model, the potential minimum is given by $V=0$ at $\phi=0$.
The potential as a function of the field value, $V(\phi)$, is shown in Ref.\cite{fr-grav}.
}
\label{fig:R2_potential}
\end{figure}

\begin{figure}[t]
\includegraphics[width=8.5cm]{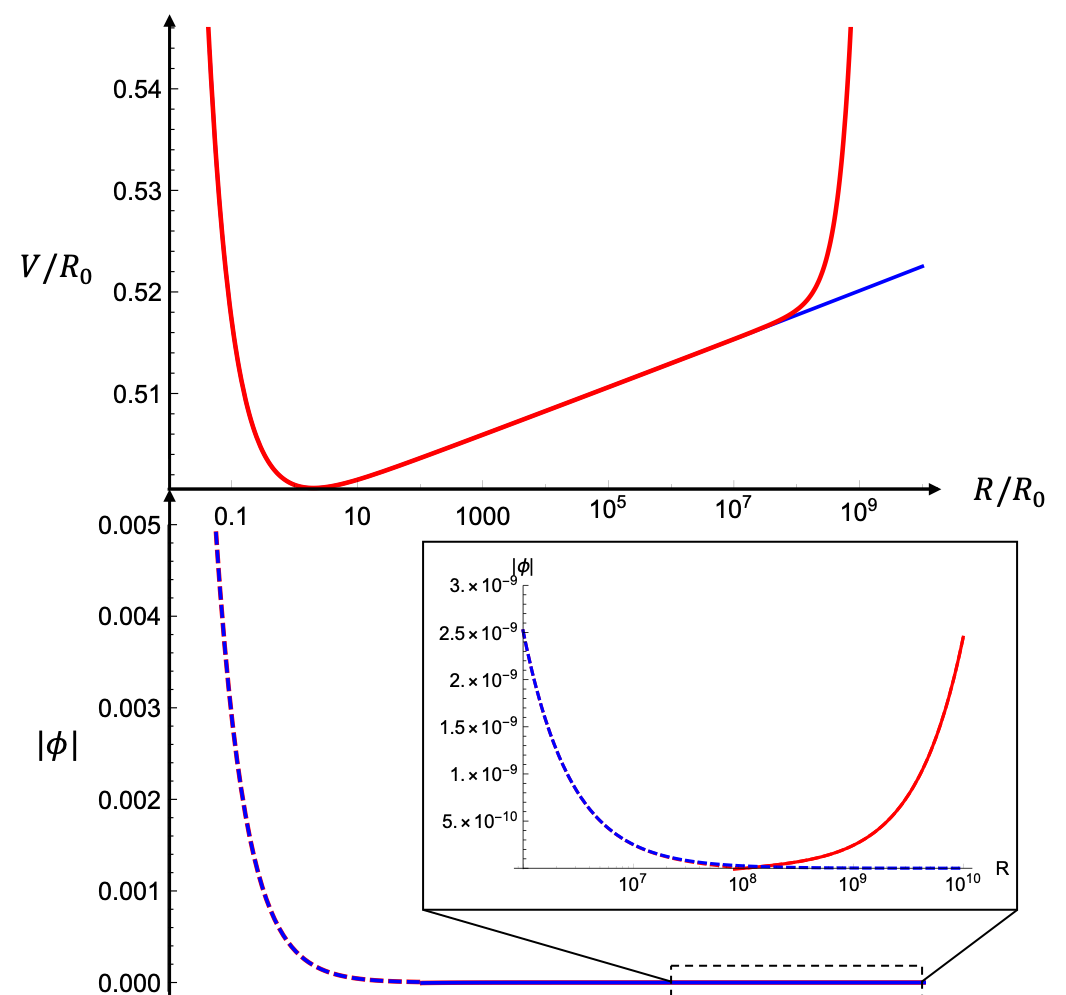}
\caption{Potential in the Einstein frame for Model 1 (red line) and power-law DE model (blue line).
We show parametric representations $V(R)$ and $\phi(R)$ in the top and the bottom, respectively.
Dashed lines show $\phi < 0$. 
The potential is positive at the minimum.
}
\label{fig:exAL-V-phi-R}
\end{figure}

The first derivative of the field potential is
\begin{align}
	V_{,\phi} &\equiv \frac{{\rm d} V}{{\rm d} \phi}  = -\frac{FR-2f}{\sqrt{6}F^2} \nonumber \\
	&= \frac{e^{-\frac{4}{\sqrt{6}}\phi} R}{\sqrt{6}}\left[ 1+\alpha (2-n) R^{n-1} - \beta (2-p) R^{p-1} \right] \,.
\end{align}
In the absence of the $\beta$ term, 
the field potential has a minimum $V=0$ at $R=0$, which corresponds to $\phi=0$ (see Fig. \ref{fig:R2_potential}).
Therefore, the expansion of the Universe is not accelerated without introducing matter.
In contrast, 
when the $\beta$ term is introduced, 
a minimum appears at $R = R_{\rm min} \simeq [\beta(2-p)]^{1/(1-p)} >0$ and $V>0$ at the minimum, 
while $R=0$ cannot be reached now because the field value $\phi$ diverges at $R = R_{F=0} \simeq [\beta p]^{1/(1-p)}>0$ (see Fig. \ref{fig:exAL-V-phi-R}).
As in the original model \cite{lalak}, 
this potential energy drives the late-time acceleration.

In the model (\ref{eq:ex lalak}), $F_{,R}$ is given by
\begin{equation}
	F_{,R} = \alpha n(n-1)R^{n-2} + \beta p(1-p) R^{p-2} \,.
\end{equation}
Thus, the second condition $F_{,R}>0$ (\ref{eq:stability conditions}) is satisfied for $n>1$, $0< p < 1$ and $\alpha, \beta>0$. 
On the other hand, the first condition $F>0$ is violated for $R < R_{F=0} \simeq [\beta p]^{1/(1-p)}$.
However, the curvature $R$ never dynamically reaches to this region 
because there is an infinite potential barrier at $R = R_{F=0}$.

\subsection{Model 2: Starobinsky DE model}

Next, we introduce the following unified model:
\begin{equation}
f(R)=R-\mu R_{\rm 0} \left[ 1- \left( 1+ \frac{R^2}{R^2_{\rm 0}} \right)^{-j} \right] + \alpha R^n,
\label{eq:Sta unified 0}
\end{equation}
where $j, \mu>0$.
For $\alpha = 0$, this model corresponds to the Starobinsky DE model \cite{StaDE}, which also satisfies the conditions (\ref{eq:stability conditions}). 
For $R \gg R_{\rm 0}$, this model can be approximated by a power-law model as
\begin{equation}
f(R) \simeq R +\mu R_{\rm 0} \left( \frac{R}{R_{\rm 0}} \right)^{-2j} + \alpha R^n\,.
\label{eq:Sta unified}
\end{equation}

\begin{figure}[t]
\includegraphics[width=8.5cm]{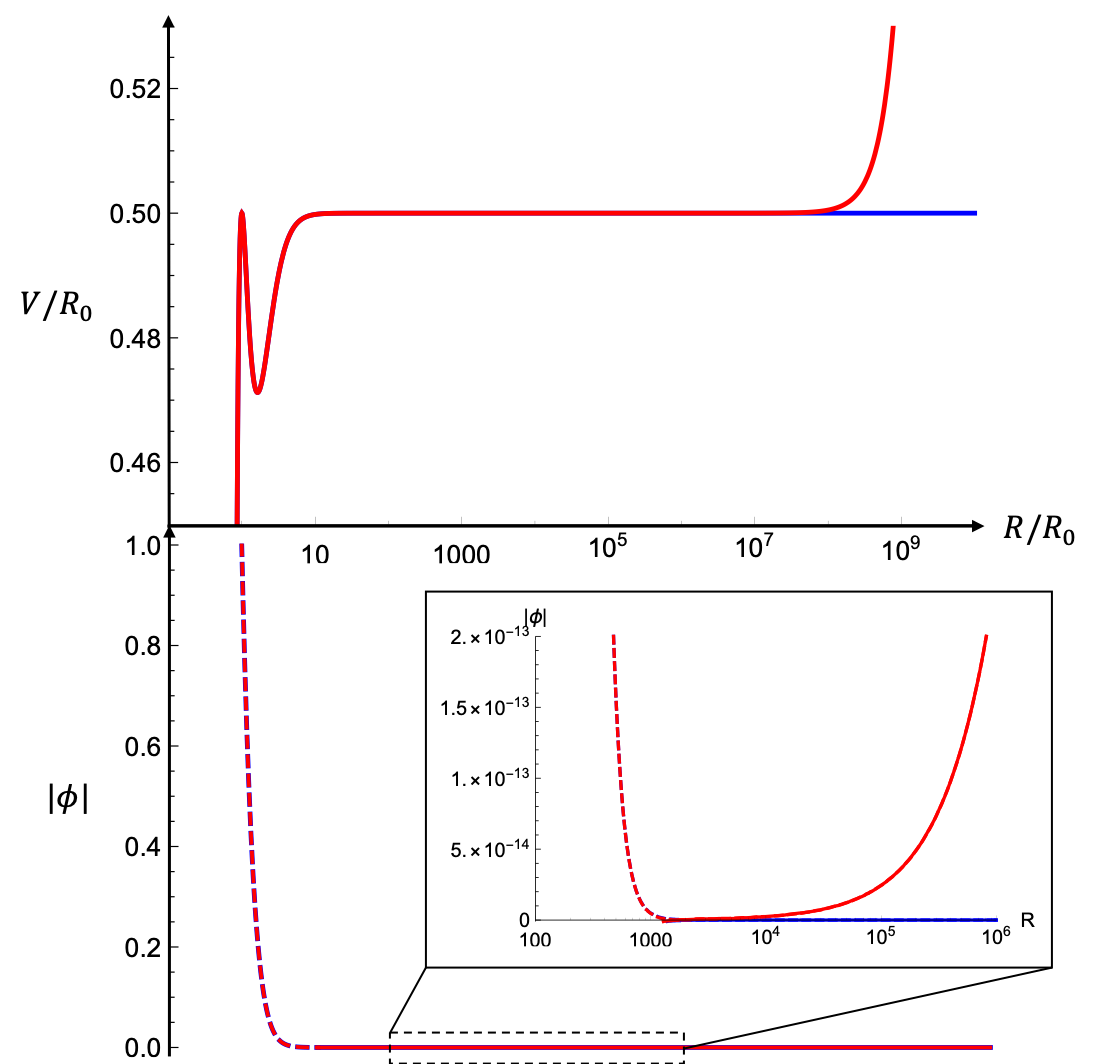}
\caption{Potential in the Einstein frame for Model 2 (red line) and Starobinsky DE model (blue line).
We show parametric representations $V(R)$ and $\phi(R)$ in the top and the bottom, respectively.
Dashed lines show $\phi < 0$. 
The potential is positive at the minimum.
}
\label{fig:V-phi_R_Sta}
\end{figure}

In the model (\ref{eq:Sta unified 0}), 
the field potential in the Einstein frame (\ref{eq:potential einstein}) and its first derivative are read as
\begin{align}
V(\phi) = &\frac{e^{-\frac{4}{\sqrt{6}}\phi}}{2} \Bigg[ \mu R_0 + \alpha (n-1) R^n - \mu R_0 \nonumber \\
&\left. \times \left( 1+\frac{R^2}{R_0^2} \right)^{-j} \left( 1 + 2j \left( 1 + \frac{R_0^2}{R^2} \right)^{-1} \right) \right] \,,
\label{eq:Sta unified V}
\end{align}
and
\begin{align}
V_{,\phi} = &\frac{e^{-\frac{4}{\sqrt{6}}\phi}}{\sqrt{6}} \Biggl[ R - 2\mu R_0 + \alpha (2-n) R^{n} \nonumber \\
&+ 2 \mu R_0 \left( 1+ \frac{R^2}{R_0^2} \right)^{-j} \left( 1 + j \left( 1 + \frac{R_0^2}{R^2} \right)^{-1} \right) \Biggr] \,,
\label{eq:Sta unified Vphi}
\end{align}
where
\begin{equation}
e^{\frac{2}{\sqrt{6}}\phi} = F(R) = 1 + \alpha n R^{n-1} - 2\mu j \frac{R}{R_0} \left(1+ \frac{R^2}{R_0^2} \right)^{-j-1}\,.
 \label{eq:phi r Sta}
  \end{equation}
 In the model (\ref{eq:Sta unified 0}), $F_{,R}$ becomes
\begin{align}
F_{,R} = &\alpha n (n-1) R^{n-2} - \frac{2 \mu j}{R_0} \left (1+ \frac{R^2}{R_0^2} \right)^{-j-1} \nonumber \\
&\times \left[ 1 - 2\mu (j+1) \left( 1 + \frac{R_0^2}{R^2} \right)^{-1} \right] \,.
\label{eq:F'_Sta_uni}
\end{align}
When the DE term is introduced, 
a minimum appears at $R=R_{\rm min} \simeq R_0$ and $V>0$ at the minimum (see Fig. \ref{fig:V-phi_R_Sta}).
Therefore, the potential energy drives the late-time acceleration in the model.

\subsection{Model 3: $g$-AB model}

\begin{figure}[t]\includegraphics[width=8cm]{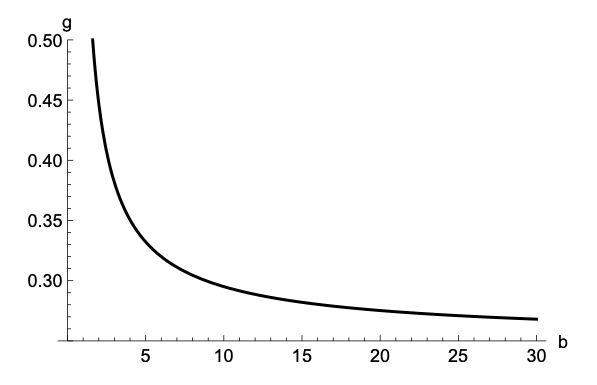}
\caption{The allowed parameter range of $g$ and $b$ in $g$-AB model.
This curve is obtained from the stability conditions (\ref{eq:stability conditions}).
}
\label{fig:g-b}
\end{figure}

Finally, we introduce the following unified model:
\begin{equation}
f(R) = (1-g)R + g \epsilon_{\rm AB} \ln \left[ \frac{\cosh (R/ \epsilon_{\rm AB} -b)}{\cosh b} \right] + \alpha R^n\,,
\label{eq:AB unified 0}
\end{equation}
where $b>0$, $0<g<1/2$, and
\begin{equation}
    \epsilon_{\rm AB} \equiv \frac{R_0}{2g\ln(1+e^{2b})}\,.
\end{equation}
This model is an extension of the model proposed by Appleby {\it et al.} \cite{ABS}, where the model has $n = 2$.
For $\alpha = 0$ and $g=1/2$, this model corresponds to the Appleby \& Battye (AB) model \cite{AB}, which also satisfies the conditions (\ref{eq:stability conditions}). 
The constraints on $g$ and $b$ are obtained from the stability conditions (\ref{eq:stability conditions}).
We show the relation between $g$ and $b$ in Fig. \ref{fig:g-b} (see \cite{ABS} for more details).
For $R \gg R_{\rm 0}$, the model (\ref{eq:AB unified 0}) can be approximated by
\begin{equation}
f(R) \simeq R - \frac{R_0}{2}+ g \epsilon_{AB} \, e^{2b} e^{-2R/ \epsilon_{AB}} + \alpha R^n\,.
\label{eq:g-AB unified}
\end{equation}

\begin{figure}[t]
\includegraphics[width=8.5cm]{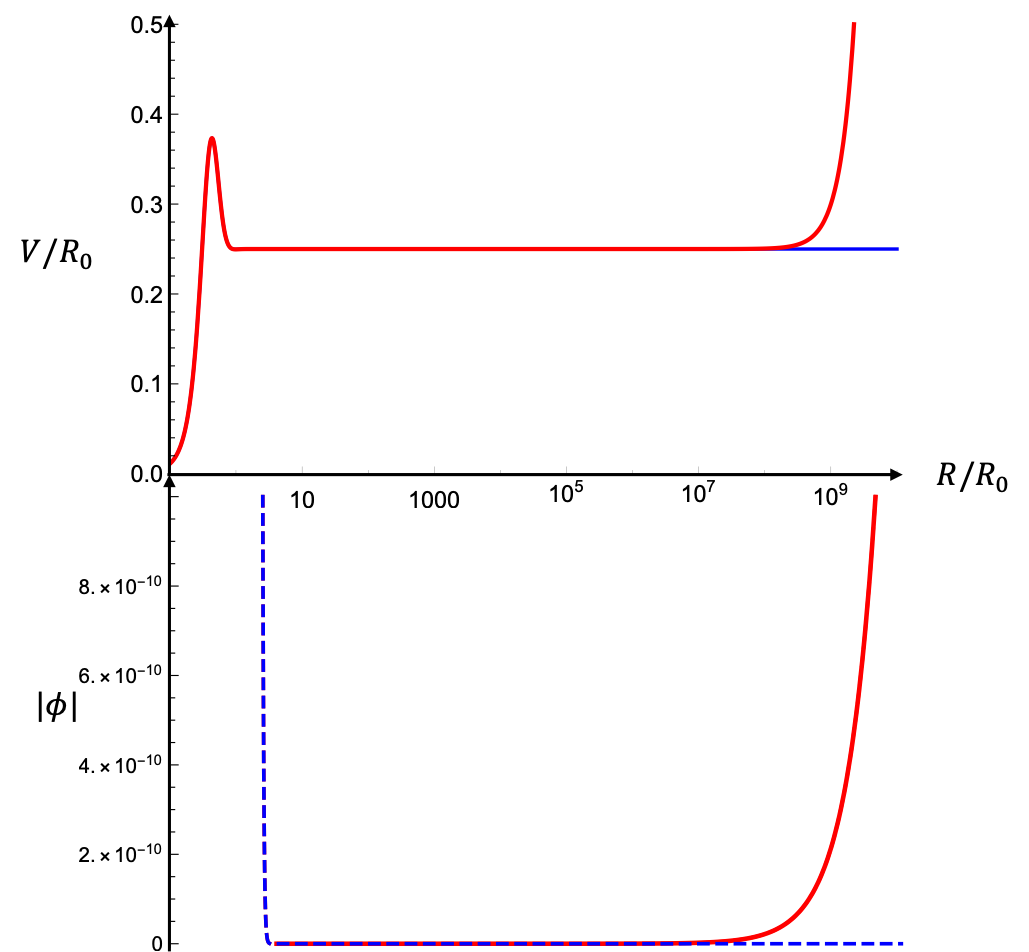}
\caption{Potential in the Einstein frame for Model 3 (red line) and AB model (blue line).
We show parametric representations $V(R)$ and $\phi(R)$ in the top and the bottom, respectively.
Dashed lines show $\phi < 0$. 
The potential is positive at the minimum.
}
\label{fig:V-phi_R_gAB}
\end{figure}

In the model (\ref{eq:AB unified 0}), 
the field potential in the Einstein frame (\ref{eq:potential einstein}) and the first derivative of the potential read,
\begin{align}
V(\phi) = &\frac{e^{-\frac{4}{\sqrt{6}}\phi}}{2} \Biggl[ \alpha (n-1) R^n +gR \tanh \left( \frac{R}{\epsilon_{\rm AB}} -b \right) \nonumber \\
&- g \epsilon_{\rm AB} \log \left( \frac{\cosh (R/ \epsilon_{\rm AB} -b)}{\cosh b}  \right) \Biggr] \,,
\label{eq:g-AB unified V}
\end{align}
and
\begin{align}
V_{,\phi} = &\frac{e^{-\frac{4}{\sqrt{6}}\phi}}{\sqrt{6}} \Biggl[ R(1-g) +\alpha (2-n) R^{n} - gR \tanh \left( \frac{R}{\epsilon_{\rm AB}} -b \right) \nonumber \\
&+ 2g \epsilon_{\rm AB} \log \left( \frac{\cosh (R/ \epsilon_{\rm AB} -b)}{\cosh b}  \right)  \Biggr] \,,
\label{eq:g-AB unified Vphi}
\end{align}
where
\begin{align}
e^{\frac{2}{\sqrt{6}}\phi} &= F(R) \nonumber \\
&= 1 - g + \alpha n R^{n-1} + g \tanh \left(\frac{R}{\epsilon_{\rm AB}} -b \right) \,.
\label{eq:phi r g-AB}
\end{align}
In the model (\ref{eq:AB unified 0}), $F_{,R}$ becomes
\begin{equation}
F_{,R} = \alpha n (n-1) R^{n-2} + \frac{g}{\epsilon_{\rm AB}} \,{\rm sech}^2 \left(\frac{R}{\epsilon_{\rm AB}} -b \right)\,.
\label{eq:F'_g-AB_uni}
\end{equation}
When the DE term is introduced, 
a minimum appears at $R = R_{\rm min} \simeq b \epsilon_{\rm AB} >0$ 
and $V>0$ at the minimum (see Fig. \ref{fig:V-phi_R_gAB}).

\section{Inflationary constraints} \label{sec:inflation}

Next, we derive an inflationary constraint on the unified models. 
Since inflation is driven without introducing any matter in $f(R)$ theories, 
we neglect the matter in this section.
In addition, the Ricci scalar $R$ is sufficiently large during inflation.
Hence, the function $f(R)$ in all the unified models above can be approximated as
\begin{equation}
f(R) \simeq R+\alpha R^n \,, 
\label{eq:a-model}
\end{equation}
which is an extension of $R^2$ model \cite{tomita,r^2} (see \cite{1311.0744} another extended $R^2$ model).
This model (\ref{eq:a-model}) corresponds to the Einstein frame action with
\begin{equation}
V(\phi)=V_0 e^{-\frac{4}{\sqrt{6}}\phi} \left[ e^{\frac{2}{\sqrt{6}}\phi} -1 \right]^{\frac{n}{n-1}} \,,
\label{eq:potentail_ab}
\end{equation}
where $V_0 \equiv \alpha(n-1)/[2(\alpha n)^{n/(n-1)}]$.

Since the comoving curvature perturbations and tensor perturbations are invariant under the conformal transformation (\ref{eq:conformal transformation}) \cite{chiba-yamaguchi, pertub_c-t}, 
the spectral index $n_s$ and the tensor-to-scalar ratio $r$ can be directly evaluated in the Einstein frame, 
where the action is equivalent to that in a single-field slow-roll inflationary model. 
Therefore, they are given as, 
\begin{equation}
r = 16\epsilon_V \,, \quad n_s = 1- 6\epsilon_V +2\eta_V \,,
\end{equation}
in terms of the slow-roll parameters,
\begin{equation}
\epsilon_V \equiv \frac{1}{2} \left( \frac{V_{,\phi}}{V} \right)^2\,, \quad  \eta_V \equiv \frac{V_{,\phi \phi}}{V} \,.
\end{equation}

In the model (\ref{eq:a-model}), the slow-roll parameters are evaluated as \cite{Motohashi},
\begin{eqnarray}
\epsilon_V &=& \frac{4E_k^2 (2-n)^2}{3[2(n-1)E_k-n]^2}, 
\label{eq:eps}\\
\eta_V &=& \frac{4(2-n)\left[ 2(2-n)E_k^2 -nE_k +n \right]}{3[2(n-1)E_k-n]^2},
\label{eq:eta}
\end{eqnarray}
where $E_k \equiv \mathrm{e}^{4(2-n)N_k/(3n)}$ and $N_k$ is e-folds from the horizon crossing to the end of inflation. 
They reduce to the well-known results,
\begin{equation}
\epsilon_V = \frac{3}{4N_k^2} \,, \quad \eta_V = -\frac{1}{N_k} \,,
\end{equation}
in the limit $n \to 2$.

We compare these predictions with the Planck data \cite{planck2018_X}:
\begin{equation}
    r < 0.01, \,\,\, n_s=0.9659 \pm 0.0041 \hspace{3mm} (95 \,\% \, {\rm CL})\,.
\end{equation}
As a result, the index $n$ in the model (\ref{eq:a-model}) is constrained as
\begin{eqnarray}
N_k = 50: 1.977< & n & <1.991 \hspace{5mm} (95 \,\% \, {\rm CL})\,, 
\label{eq:constraint of n 50} \\
N_k = 60: 1.991 < & n & < 2.003 \hspace{5mm} (95 \,\% \, {\rm CL})\,.
\label{eq:constraint of n}
\end{eqnarray}
Note that this constraint is slightly relaxed from our previous result $1.965< n<2$ \cite{my1st}.
In the original model $f(R)=R+\alpha R^n - \beta R^{2-n}$, $p$ is fixed to be $p=2-n$. Therefore, the viable condition $0<p<1$ restricts $n$ as $1<n<2$.
In the extended model (\ref{eq:ex lalak}), this constraint is absent.

In addition, we can estimate the model parameter $\alpha$ from the amplitude of the scalar power spectrum,
\begin{equation}
{\cal P_R} = \frac{H_E^2}{8\pi^2 \epsilon_V} \simeq \frac{N_k^2}{144\pi^2 \alpha} \,,
\label{eq:curvature pertub}
\end{equation}
where $H_E$ is the Hubble parameter in the Einstein frame and we used $n \simeq 2$ in the second equation.
Comparing this with the CMB observation  ${\cal P_R} \sim 10^{-10}$ \cite{planck2018_X},
we find
\begin{equation}
\alpha \sim 10^{10}  \,,
\label{eq:alpha constraint}
\end{equation}
for $N_k=50 \-- 60$.\\

  \section{local-gravity constraint}
\label{sec:gravity test}

Here, we consider the local-gravity constraint on the unified models (\ref{eq:ex lalak}), (\ref{eq:Sta unified}) and (\ref{eq:g-AB unified})
following the analysis in Refs. \cite{fr-DE, ABS}.
In this section, we work in the Einstein frame, 
where it is more clear that the extra degree of freedom, i.e. scalaron, mediates the fifth force as
\begin{equation}
\vec{F}_\phi = \frac{M}{\sqrt{6}} \vec{\nabla}\phi.
\label{eq:fifth force}
\end{equation}

In this analysis, we study the configuration of the scalar field in the finite density region, 
which is governed by Eq. (\ref{eq:phi eom cov}):
\begin{equation}
\widetilde{\square} \phi = \frac{{\rm d} V_{\rm eff}}{{\rm d} \phi} \,; \quad V_{\rm eff} (\phi)= V(\phi) + \rho e^{-\frac{\phi}{\sqrt{6}}} \,.
\end{equation}

The minimum of the effective potential $\phi = \phi_{\rm min}$ depends on the density $\rho$ through
\begin{equation}
    \frac{{\rm d} V_{\rm eff}}{{\rm d} \phi} = V_{,\phi} - \frac{\rho e^{-\frac{\phi_{\rm min}}{\sqrt{6}}}}{\sqrt{6}} = 0 \,,
\label{eq:scalaron mass}
\end{equation}
and the scalaron's mass also depends on $\rho$ through
\begin{equation}
    m_{\phi}^2 \equiv \frac{d^2 V_{\rm eff}(\phi_{\rm min})}{d\phi^2} = V_{,\phi\phi} + \frac{\rho e^{-\frac{\phi_{\rm min}}{\sqrt{6}}}}{6},
\label{eq:scalaron mass}
\end{equation}
at the minimum $\phi = \phi_{\rm min}$. 
These imply that the range of the fifth force changes according to the density in its environment.
This is known as the chameleon screening mechanism \cite{chameleon,chameleon2}.
In the following, 
we see if the unified models (\ref{eq:ex lalak}), (\ref{eq:Sta unified}) and (\ref{eq:g-AB unified}) can evade the local-gravity constraint by the chameleon screening mechanism.

\subsection{\label{sec:chameleon}Chameleon screening mechanism}
First, we briefly review the chameleon screening mechanism for the fifth force sourced by a star, 
modeling it as a spherically symmetric non-relativistic object with a constant density $\rho_c$. 
It is assumed that the star is surrounded by baryons and dark matter with a homogeneous density $\rho_{\rm G} \simeq 10^{-24} {\rm g/cm^3}$ \cite{chameleon}.
In this analysis, we focus on the vicinity of the star and neglect the cosmic expansion.

Assuming a static and spherically symmetric profile, 
the field equation reduces to
\begin{equation}
\frac{d^2\phi}{d\tilde{r}^2}+\frac{2}{\tilde{r}} \frac{d\phi}{d\tilde{r}} = \frac{dV_{\rm eff}}{d\phi} \,.
\label{eq:phi eom}
\end{equation}
In the outer region, 
the scalaron relaxes to the minimum $\phi_G \equiv \phi_{\rm min}(\rho_G)$ as
\begin{equation}
\phi \simeq \phi_G + \frac{Qe^{-m_G \tilde{r}}}{\tilde{r}} \,,
\end{equation}
where $m_G \equiv m_\phi(\rho_G)$. 
Here, $Q$ is the scalar charge and determined by matching to an inner solution.
Inside the star, the effective potential has the minimum at $\phi_c \equiv \phi_{\rm min}(\rho_c)$ with a mass $m_c \equiv m_\phi(\rho_c)$. 
When the Compton wavelength $\lambda_c \sim 1/m_c$ is much shorter than the radius of the star $\tilde r_c$, 
\begin{equation}
	\lambda_c \ll \tilde r_c \quad (m_c \tilde r_c \gg 1) \,,
	\label{eq:thin shell condition}
\end{equation}
the scalaron stays near $\phi=\phi_c$ in the inside of the star and has a thin-shell profile.
In this case,  
the scalar charge is given by \cite{chameleon},
\begin{equation}
	\frac{Q}{M_c} = \frac{3}{4\sqrt{6}\pi} \left( \frac{\Delta \tilde r_c}{\tilde r_c} \right) \,,
\end{equation}
in the unit of the mass of the star $M_c$. 
Here, the thin-shell parameter $\Delta \tilde r_c/\tilde r_c$ is given as,
\begin{equation}
	\frac{\Delta \tilde r_c}{\tilde r_c} = \frac{\phi_G - \phi_c }{\sqrt{6} \Phi_c} \,; \quad \Phi_c \equiv \frac{M_c}{8\pi \tilde r_c} \,.
	\label{eq:thin shell parameter}
\end{equation}
Therefore, the fifth force (\ref{eq:fifth force}) is screened when the thin-shell parameter is small enough.\\

When the inflationary term is absent, 
the $\phi_G$ term is dominant and 
the models (\ref{eq:ex lalak}), (\ref{eq:Sta unified}) and (\ref{eq:g-AB unified}) approximately give,
    \begin{align}
        \left|\frac{\Delta \tilde r_c}{\tilde r_c}\right| \simeq 
        \begin{cases}
            \frac{2 \tilde \beta p}{\Phi_c} \left(\frac{\rho_{G}}{R_0}\right)^{p-1} & \text{Power-law model (\ref{eq:ex lalak})} \\
            \frac{4 \mu j}{\Phi_c} \left(\frac{\rho_{G}}{R_0}\right)^{-2j-1} & \text{Starobinsky DE model (\ref{eq:Sta unified})} \\
            \frac{g e^{2b}}{\Phi_c} \exp\left(-\frac{2\rho_{G}}{\epsilon_{AB}}\right) & \text{$g$-AB model (\ref{eq:g-AB unified})}
        \end{cases}
        \label{eq:thin shell}
    \end{align}
where we have introduced the dimensionless parameter,
    \begin{equation}
        \tilde \beta \equiv \beta R_0^{p-1} \,.
        \label{eq:tilde beta}
    \end{equation} 
Here, the parameters $\tilde{\beta}$ and $\mu$ are estimated to be order of unity to explain the present value of the dark energy density parameter, $\Omega_{\rm DE} \simeq 0.7$.
In addition, from Fig. \ref{fig:g-b}, $b>0$ under the constraint on $g$. 
From these expressions, 
one can see that the thin-shell parameter quickly decreases as $\rho_G$ increases in the latter two models but only slowly in the first model. 
Reflecting this fact, the power-law model is tightly constrained by local-gravity tests while the other two are viable.
In Table \ref{tab:const for DE}, 
we summarize the solar-system constraints for the model parameters in these DE models \cite{capo_tsujikawa, ABS}. 
Here, the constraint is not shown for the $g$-AB model 
because the solar-system constraint does not restrict the parameter region 
under the stability conditions (see Fig. \ref{fig:g-b}).
\begin{table}[ht]
\caption{The solar-system constraints on the model parameters \cite{capo_tsujikawa, ABS}.}
\label{tab:const for DE}
\centering
    \begin{tabular}{|l|c|c|c|} \hline
      DE models & constraint \\ \hline\hline
      Power-law model & $p<10^{-10}$\\ \hline
      Starobinsky DE model & $j>0.9$ \\ \hline
      $g$-AB model & \-- \\ \hline
    \end{tabular}
\vspace{0.5cm}
\end{table}

\subsection{\label{sec:mass} Constraints in the unified models}

As Figs. \ref{fig:exAL-V-phi-R}-\ref{fig:V-phi_R_gAB} indicate, 
the inflationary term can affect the scalaron's potential even for curvatures lower than the inflationary scale. 
Here, therefore, 
we see how the local-gravity constraints can change when the inflationary term $\alpha R^n$ is added. 
For later convenience, we make $\alpha$ dimensionless as
    \begin{equation}
        \tilde \alpha = \alpha R_0^{n-1} \,,
        \label{eq:tilde alpha}
    \end{equation}
which is estimated to be
\footnote{Strictly speaking, this value depends on $n$ but the deviation is irrelevant under the constraint in Sec. \ref{sec:inflation}, $n-2 < {\cal O}(10^{-2})$.}
\begin{equation}
	\tilde \alpha \sim 10^{-110} \,.
\end{equation}
from the normalization (\ref{eq:alpha constraint}).

\subsubsection{Model 1: Power-law DE + $R+\alpha R^n$ model}

First, we consider the model (\ref{eq:ex lalak}):
\begin{equation}
f(R) = R+ \tilde\alpha R_0 \left(\frac{R}{R_0}\right)^n - \tilde\beta R_0 \left(\frac{R}{R_0}\right)^p \,,
\end{equation}
where the parameters are made dimensionless as in Eqs. (\ref{eq:tilde beta}) and (\ref{eq:tilde alpha}). 
From Table \ref{tab:const for DE}, this model is tightly constrained by the solar-system observations. 
Here, we verify whether the inflationary term can evade this constraint.

The minimum of the effective potential in this model is determined by
\begin{equation}
	\frac{R}{R_0} + \tilde{\alpha} (2-n) \left( \frac{R}{R_0} \right)^n - \tilde{\beta} (2-p) \left(\frac{R}{R_0}\right)^p = \frac{\rho e^{-\frac{\phi}{\sqrt{6}}} }{R_0} \,.
	\label{eq:min eff}
\end{equation}
with
\begin{equation}
 e^{\frac{2}{\sqrt{6}}\phi} = F(R) = 1 + \tilde \alpha n \left( \frac{R}{R_0} \right)^{n-1} - \tilde \beta p \left( \frac{R}{R_0} \right)^{p-1} \,.
 \label{eq:phi r 2}
\end{equation}
As mentioned earlier, the coefficients are roughly estimated to be, 
\begin{equation}
	\tilde \alpha \sim 10^{-110} \,, \quad \tilde \beta \sim 1 \,.
	\label{eq:alphabeta}
\end{equation}
On the other hand, for $\rho = {\cal O}(10^{-24} \-- 1) {\rm g/cm^3}$, 
the right-hand side is roughly estimated to be,
\begin{equation}
	\frac{\rho e^{-\frac{\phi}{\sqrt{6}}} }{R_0} \sim 10^4 \-- 10^{28} \,,
\end{equation}
where we have used $R_0 \simeq 12H_0^2 = 4\rho_{\rm crit} \sim 10^{-29} {\rm g/cm^3}$.
Therefore, at the minimum in the relevant region, 
the first term in Eq. (\ref{eq:min eff}) is dominant and the curvature scale is given by
\begin{equation}
	\frac{R_{\rm min}}{R_0} \simeq \frac{\rho}{R_0} \sim 10^{28} \left(\frac{\rho}{ 1 {\rm g/cm^3}} \right) \,.
	\label{eq:rscale}
\end{equation}
The corresponding field value at the minimum is determined by Eq. (\ref{eq:phi r 2}). 
For the typical values (\ref{eq:alphabeta}) and (\ref{eq:rscale}), $\phi_{\rm min} \ll 1$ and hence
\begin{equation}
 	\frac{2\phi_{\rm min}}{\sqrt{6}} \simeq \tilde \alpha n \left( \frac{R_{\rm min}}{R_0} \right)^{n-1} - \tilde \beta p \left( \frac{R_{\rm min}}{R_0} \right)^{p-1} \,.
\end{equation}
Therefore, the modified terms are important to determine the field value $\phi_{\rm min}$.
Among these two terms, the $\alpha$ term is negligible 
unless the index $p$ is extremely small,
\begin{equation}
	p \ll \frac{n \tilde \alpha}{\tilde \beta} \left( \frac{R_{\rm min}}{R_0} \right)^{n-p} \sim 10^{-54-28p} \left(\frac{\rho}{ 1 {\rm g/cm^3}} \right)^{n-p} \,,
\end{equation}
or unless the local density is sufficiently large,
\begin{equation}
	 \rho \gg \left(\frac{p \tilde \beta}{n \tilde \alpha}\right)^{\frac{1}{n-p}} R_0 \sim \left(\rho_{\inf}^{n-1}\rho_{\rm crit}^{1-p} \right)^{\frac{1}{n-p}} \,,
\label{eq:rho alpha effect}
\end{equation}
where $\rho_{\rm inf}$ is the energy density of inflation, $\rho_{\rm inf} \simeq R_{\rm inf} \equiv \tilde{\alpha}^{-1/(n-1)}R_0$.
In this case, the field value is given in terms of the density as
\begin{equation}
 	\frac{2\phi_{\rm min}}{\sqrt{6}} \simeq - \tilde \beta p \left( \frac{R_{\rm min}}{R_0} \right)^{p-1} \simeq - \tilde \beta p \left( \frac{\rho}{R_0} \right)^{p-1} \,.
\end{equation}

We can find that a similar approximation can be applied to the effective mass. 
Under this approximation, the effective mass is estimated to be,
\begin{align}
 	m_\phi^2 \simeq R_0 \left[\frac{1}{3p(1-p)\tilde \beta}\left(\frac{\rho}{R_0}\right)^{2-p} - \frac{5\rho}{6R_0}\right] \,,
\end{align}
For small values of $p$, 
the first term gives the dominant contribution,
\begin{align}
 	m_\phi^2 &\simeq \frac{R_0}{3p(1-p)\tilde \beta}\left(\frac{\rho}{R_0}\right)^{2-p} \nonumber \\
 	 &\sim \frac{H_0^2}{p}\left(\frac{\rho}{\rho_{\rm crit}}\right)^{2-p} \,.
\end{align}
and the thin-shell parameter is unchanged from Eq. (\ref{eq:thin shell}).
From the arguments above, we found that the $\alpha$ term is irrelevant to the local-gravity analysis under the inflationary and DE constraints. 
Therefore, the local-gravity constraints in the model (\ref{eq:ex lalak}) reduce to those for the DE model \cite{plmodel},
\begin{equation}
	f(R) = R - \beta R^{p} \,.
\end{equation}
In Ref. \cite{lalak}, where the model with $p=2-n$ is treated, 
they claimed without an analysis on the fifth force that their unified model is consistent with the local-gravity tests. 
Contrary to this statement, 
the power-law unified model is not viable for the local-gravity tests.

\subsubsection{Model 2: Starobinsky DE + $\alpha R^n$ model}

The above argument also showed that, in the model (\ref{eq:ex lalak}), the inflationary term can affect the local-gravity constraints for the density,
\begin{equation}
	 \rho \gg \rho_{\rm th} \equiv \left(\rho_{\inf}^{n-1}\rho_{\rm crit}^{1-p} \right)^{\frac{1}{n-p}} \,,
\end{equation}
which is higher than the density relevant to the solar-system constraints but can be much lower than the inflationary scale. 
This observation indicates that 
the inflationary term can affect the local-gravity analysis depending on the model parameters in general unified models. 
Next, 
we see how the effect of the inflationary term appears in the viable models (\ref{eq:Sta unified 0}) and (\ref{eq:g-AB unified}).

First, we consider the model (\ref{eq:Sta unified 0}). 
In a high density region, it can be approximated by the power-law model (\ref{eq:Sta unified}):
\begin{equation}
f(R) = R +\mu R_{\rm 0} \left( \frac{R}{R_{\rm 0}} \right)^{-2j} + \tilde \alpha R_0 \left(\frac{R}{R_0}\right)^n\,.
\end{equation}
Therefore, the analysis is parallel to the model (\ref{eq:ex lalak}) with the replacements $\tilde \beta \to -\mu$ and $p \to -2j~(j>0.9)$, i.e. a possible value of the index is different.
Taking into this fact, 
the threshold density can be estimated as
\begin{equation}
	 \rho_{\rm th} \equiv \left(\rho_{\inf}^{n-1}\rho_{\rm crit}^{1+2j} \right)^{\frac{1}{n+2j}} \,.
\end{equation}
This shows that the inflationary term can be important even for a very low value of the density when $j$ is large.

We estimate the effective mass,
\begin{equation}
    m_\phi^2 = \left. \frac{1}{6}\left( \frac{FR-6f}{F^2} + \frac{2}{F_{,R}} \right) \right|_{R=R_{\rm min}}
    \label{eq:phi mass R}
\end{equation}
without ignoring the $\alpha$ term.
The first term is dominated by the GR term and estimated to be,
\begin{equation}
    \left. \frac{FR-6f}{6F^2} \right|_{R=R_{\rm min}} \simeq -\frac{5R_{\rm min}}{6} \simeq -\frac{5\rho}{6} \,.
\end{equation}
The second term $1/(3F_{,R})$ is determined by the inflationary/DE terms and larger than the first term for a large value of the density. 
Therefore, in terms of the Compton wavelength $\lambda_\phi = m_{\phi}^{-1}$, 
it is given by a sum of the inflationary and DE pieces as
\begin{align}
    \lambda_{\phi}^2 &\simeq 3F_{,R} \nonumber \\
    &= \frac{n(n-1)}{H_{\rm inf}^2} \left( \frac{\rho}{\rho_{\rm inf}} \right)^{n-2} + \frac{\mu j(2j+1)}{2H_0^2} \left( \frac{\rho}{4\rho_{\rm crit}} \right)^{-2j-2} \nonumber  \\
    &\sim j^2 H_0^{-2} \left( \frac{\rho}{\rho_{\rm crit}} \right)^{-2j-2}\left[ 1 + {\cal O}(1)\left(\frac{\rho}{\rho_{\rm th}}\right)^{2j+n}\right] \,,
    \label{eq:lambda_Sta}
\end{align}
where we have introduced the inflationary Hubble scale $H_{\rm inf}^2 \equiv \rho_{\rm inf}/3$.

\begin{figure}[t]
	\centering
		\includegraphics[width=8cm]{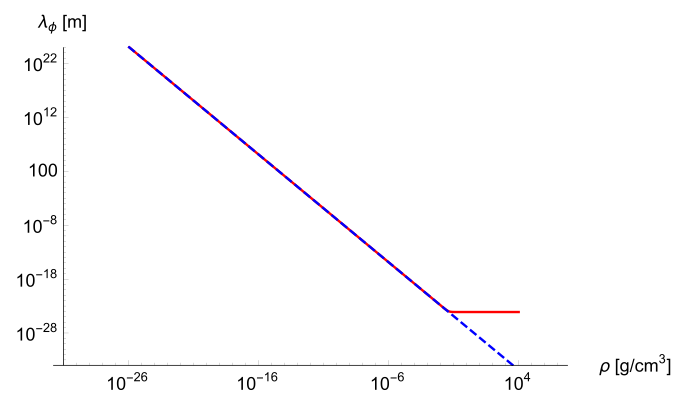}
	\caption{The relation between the Compton wavelength $\lambda_\phi = m_\phi^{-1}$ and the density  $\rho$. Blue dashed line and red solid line correspond to the DE model and unified model (\ref{eq:Sta unified}), respectively. In this figure, we set $j=1$ and $n=2$.
}
	\label{fig:Sta-mass_n1}
\end{figure}

\begin{figure}[t]
	\centering
		\includegraphics[width=8cm]{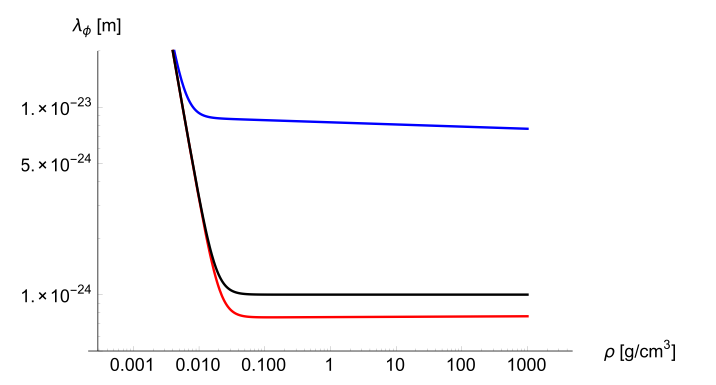}
	\caption{The relation between the Compton wavelength $\lambda_\phi = m_\phi^{-1}$ and the density $\rho$ for possible values of $n$ under the inflationary constraint (\ref{eq:constraint of n}). 
	These lines correspond to $n=1.977, 2$ and $2.003$ from bottom to top.
	In this figure, we fix $\rho_{\rm inf}$ in Eq. (\ref{eq:lambda_Sta}).
	}
	\label{fig:Sta_lambda_neq2}
\end{figure}

This result shows that the Compton wavelength first rapidly decreases as the density increases but becomes approximately constant for $\rho > \rho_{\rm th}$ with $n \simeq 2$ (see Figs. \ref{fig:Sta-mass_n1} \-- \ref{fig:Sta_lambda_neq2}). 
In addition, the inflationary term can dominate over the DE term at smaller scales than the Earth's scale for $j>0.9$. 
The Compton wavelength can be enhanced by the factor $(\rho/\rho_{\rm inf})^{n-2}$ for $n<2$. 
However, it is not large enough under the inflationary constraint (\ref{eq:constraint of n}). 
Therefore, the thin-shell condition (\ref{eq:thin shell condition}) is kept even when the inflationary term is added. 
\footnote{On the other hand, from the viewpoint of the inflationary model, the DE term makes the scalaron's mass light in a low dense region and fifth-force constraints relevant.}

\begin{figure}[t]
		\includegraphics[width=8cm]{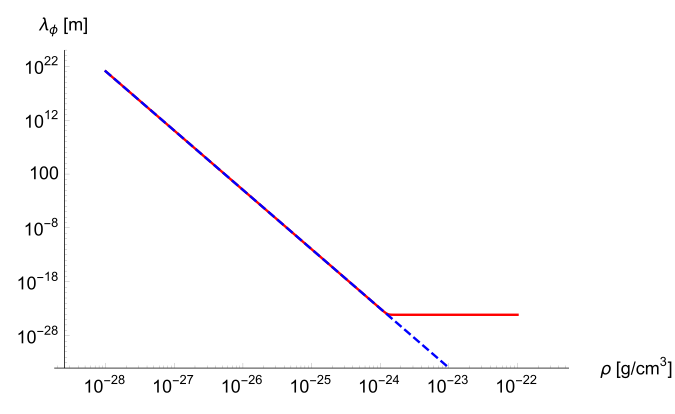}
	\caption{Same as Fig. \ref{fig:Sta-mass_n1} except for $j=10$.}
	\label{fig:Sta-mass_n10}
\end{figure}

Then, let us see next how the thin-shell parameter (\ref{eq:thin shell parameter}) is modified. 
With the inflationary term, the scalar field is related to the curvature/density as
\begin{align}
 	\frac{2\phi_{\rm min}}{\sqrt{6}} &\simeq \tilde \alpha n \left( \frac{R_{\rm min}}{R_0} \right)^{n-1} - 2\mu j \left( \frac{R_{\rm min}}{R_0} \right)^{-2j-1} \nonumber \\
 	&\sim \left( \frac{\rho}{\rho_{\rm crit}} \right)^{-2j-1}\left[ 1 + {\cal O}(1)\left(\frac{\rho}{\rho_{\rm th}}\right)^{2j+n} \right] \,.
 	\label{eq:phi min}
\end{align}
This shows that $|\phi_{\rm min}|$ increases as $\rho$ increases for $\rho > \rho_{\rm th}$ in contrast that it decreases when the inflationary term is absent. 
Hence, the $\phi_c$ term can be also important in Eq. (\ref{eq:thin shell parameter}).

In analyzing the thin-shell parameter (\ref{eq:thin shell parameter}) from the field value (\ref{eq:phi min}), 
there are three possibilities: 
(i) the DE term dominates over all regions $\rho_{\rm th} > \rho_c$, (ii) the inflationary term dominates over all regions $\rho_{\rm th} < \rho_G$, and (iii) the inflationary term dominates in the inner region of the object while the DE term does in the outer region $\rho_G < \rho_{\rm th} < \rho_c $. 
In the first case, the analysis reduces to that for the DE model. 
In the second case, the scalaron is also heavy in the outer region and hence it is not necessary to discuss the screening mechanism. 
Therefore, the relevant case is $\rho_G < \rho_{\rm th} < \rho_c $. 
In this case, the thin-shell parameter is estimated to be
    \begin{align}
        \left| \frac{\Delta \tilde r_c}{\tilde r_c} \right| \simeq \frac{2}{\Phi_c}\left[ 2\mu j \left(\frac{\rho_{G}}{R_0}\right)^{-2j-1} + \tilde\alpha n \left( \frac{\rho_c}{R_0} \right)^{n-1} \right]\,,
    \end{align}
where now the $\rho_c$-dependent term can be non-negligible. 
The additional term is roughly estimated to be
\begin{align}
        \frac{2\tilde\alpha n}{\Phi_c}\left( \frac{\rho_c}{R_0} \right)^{n-1} \sim \frac{1}{\Phi_c}\left( \frac{\rho_c}{\rho_{\rm inf}} \right)^{n-1} \sim \frac{1}{H_{\rm inf}^2 \tilde r_c^2}\left( \frac{\rho_c}{\rho_{\rm inf}} \right)^{n-2} \,,
\end{align}
where we have estimated the gravitational potential as $\Phi_c \sim \rho_c \tilde r_c^2$. 
For a fixed size of the object $\tilde r_c$, the parameter can be enhanced by the factor $(\rho_c/\rho_{\rm inf})^{n-2}$ for $n<2$. 
However, it is not large enough under the inflationary constraint (\ref{eq:constraint of n}).

\subsubsection{Model 3: $gR^n$-AB model}

\begin{figure}[t]
\centering
		\includegraphics[width=8cm]{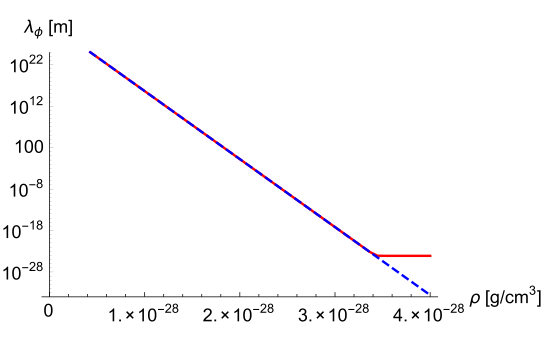}
	\caption{The relation between the Compton wavelength $\lambda_\phi = m_\phi^{-1}$ and the density  $\rho$. Blue dashed line and red solid line correspond to the DE model and unified model (\ref{eq:g-AB unified}), respectively. In this figure, we set $b=2, g=0.47$ and $n=2$.
}
	\label{fig:g-AB-lambda_g047_b2}
\end{figure}

\begin{figure}[t]
		\includegraphics[width=8cm]{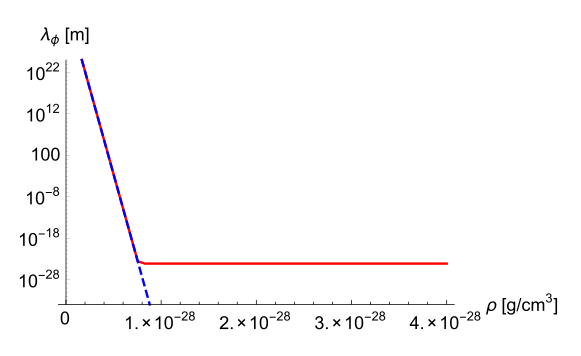}
	\caption{Same as Fig. \ref{fig:g-AB-lambda_g047_b2} except for $b=17, g=0.28$.}
	\label{fig:g-AB-lambda_g028_b17}
\end{figure}

Finally, we consider the model (\ref{eq:AB unified 0}). 
In a high density region, 
it can be approximated by the exponential model (\ref{eq:g-AB unified})
\begin{equation}
    f(R) \simeq R - \frac{R_0}{2}+ g \epsilon_{AB} \, e^{2b} e^{-2R/ \epsilon_{AB}} + \tilde \alpha R_0 \left(\frac{R}{R_0}\right)^n\,.
\end{equation}
As in the other models, the curvature scale at the potential minimum can be estimated as
\begin{equation}
    \frac{R_{\rm min}}{R_0} \simeq \frac{\rho}{R_0} \,.
\end{equation}
The corresponding field value is
\begin{align}
 	\frac{2\phi_{\rm min}}{\sqrt{6}} &\simeq \tilde\alpha n \left( \frac{R_{\rm min}}{R_0} \right)^{n-1} - 2g e^{2b} e^{-2R_{\rm min}/ \epsilon_{AB}} \nonumber \\
 	&= n \left( \frac{\rho}{\rho_{\rm inf}} \right)^{n-1} - 2g e^{2b} e^{-2\rho/ \epsilon_{AB}} \,,
\end{align}
and the Compton wavelength is
\begin{align}
 	\lambda_\phi^2 \simeq \frac{n(n-1)}{H_{\rm inf}^2} \left( \frac{\rho}{\rho_{\rm inf}} \right)^{n-2} + \frac{2g^2 e^{2b}\ln(1+e^{2b})}{H_0^2} e^{-2\rho/ \epsilon_{AB}} \,.
\end{align}
We show the relation between the Compton wavelength $\lambda_\phi = m_\phi^{-1}$ and the density $\rho$ with $n=2$ in Figs. \ref{fig:g-AB-lambda_g047_b2} \-- \ref{fig:g-AB-lambda_g028_b17}. 
The DE term decreases and the inflationary term becomes relevant for a lower value of the density than the other models. 
However, 
the Compton wavelength is asymptotic to the same small value as in the other models. 
Therefore, the thin-shell condition (\ref{eq:thin shell condition}) is kept even when the inflationary term is added.

\begin{figure}[t]
		\includegraphics[width=8cm]{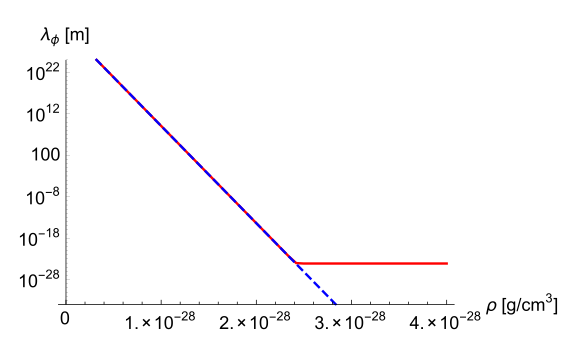}
	\caption{Same as Fig. \ref{fig:g-AB-lambda_g047_b2} except for $b=3, g=0.45$.}
	\label{fig:g-AB-lambda_g045_b3}
\end{figure}
\begin{figure}[t]
		\includegraphics[width=8cm]{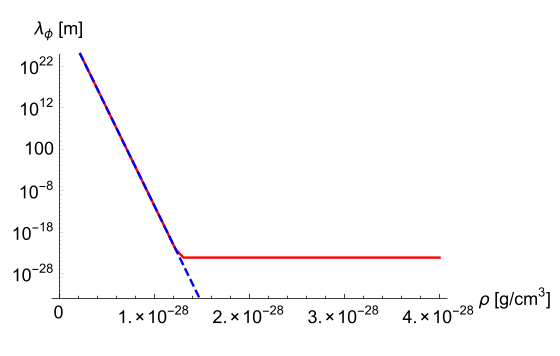}
	\caption{Same as Fig. \ref{fig:g-AB-lambda_g047_b2} except for $b=9, g=0.3$.}
	\label{fig:g-AB-lambda_g03_b9}
\end{figure}

The thin-shell parameter is estimated to be
    \begin{align}
        \left| \frac{\Delta \tilde r_c}{\tilde r_c} \right| \simeq \frac{1}{\Phi_c}\left[ g e^{2b} \exp\left(-\frac{2\rho_{G}}{\epsilon_{AB}}\right) + 2 \tilde\alpha n \left( \frac{\rho_c}{R_0} \right)^{n-1} \right]\,.
    \end{align}
The correction from the inflationary term never becomes large under the inflationary constraint (\ref{eq:constraint of n}).

\section{Conclusion}\label{sec:conclusion}

In this paper, we studied cosmological and local-gravity tests on unified models of inflation and dark energy in $f(R)$ gravity for three unified models: the power-law DE model, the Starobinsky DE model, and the $g$-AB DE model with the inflationary $R^n$ term. 

From the observation of the primordial fluctuations by the Planck satellite, 
we have obtained a constraint on the index $n$ and found that it should close to two: $|n-2|<{\cal O}(0.01)$.
Moreover, the amplitude of the fluctuations determines the scale of the inflationary term.

Next, we studied the local-gravity test in the unified models.
In contrast to a naive expectation, 
we found that the inflationary term can be relevant to the analysis even for a density much lower than the inflationary scale. 
Then, we reanalyzed the local-gravity tests of each DE model by carefully incorporating the inflationary term. 

First, the power-law DE model has been tightly constrained by the solar-system observations. 
Therefore, the main concern in this model is whether the inflationary term can affect the local-gravity analysis for the densities in the solar-system. 
As a result, we found that the threshold density cannot be low enough to affect the analysis. Thus, this unified model is also tightly constrained.

Second, the other two models have a large viable parameter region as DE models. 
Therefore, for these models,  we studied whether the fifth force is still well screened for objects with various values of the density in the unified models.
In these models, the threshold density can be very low and then the inflationary term dominates in the object while the DE term does in the environment. We reanalyse the local gravity constraints in this case and found that the corrections to the scalaron's mass and thin-shell parameter are negligible under the inflationary constraint on the index $n$.

In conclusion, 
while the inflationary term is non-negligible in the analysis, 
it does not affect the local-gravity constraints at a detectable level for the three unified models treated in this paper. 
However, 
it would be remarkable that the large hierarchy between the inflationary and astrophysical scales is not sufficient to show this conclusion; we have used the inflationary constraints (\ref{eq:constraint of n 50}) or (\ref{eq:constraint of n}). 
For example, the higher curvature term $R^n$ with $n = 1.2$ can give a few percent modification on the thin-shell parameter. 
Therefore, our analysis would give non-trivial implications on the construction of more general inflationary models or, not restricted to the inflationary one, possiblely higher curvature terms in a model.

\section*{Acknowledgements}

This research was supported in part by the JSPS Grant-in-Aid for Scientific Research
(M.Y: Grant No.19J12990, R.S: Grants No. 17K14286, No. 19H01891).

\end{document}